\documentclass[12pt]{iopart}

\usepackage{cite}
\usepackage[usenames,dvipsnames]{color}
\usepackage{iopams}
\usepackage{placeins}
\usepackage{graphics}
\usepackage{epsfig}
\usepackage{multirow}

\def\etal\;{{\it et al.}}

\font\bigastfont=cmr10 scaled \magstep 1
\def\bdot{\hbox{\bigastfont .}}

\newcommand{\CD}{{\cal D}}
\newcommand{\CE}{{\cal E}}

\newcommand{\CQ}{{Q}}
\newcommand{\CR}{{\cal R}}

\newcommand{\CW}{{\cal W}}

\newcommand{\average}[1]{\left\langle #1 \right\rangle_\CD}
\newcommand{\initial}[1]{{#1_{\it i}}}

\begin{document}

\title[Global gravitational instability of FLRW backgrounds]{Global~gravitational~instability~of~FLRW~backgrounds\\ ---interpreting the dark sectors}

\author{Xavier Roy$^{1}$, Thomas Buchert$^{1}$, Sante Carloni$^{2}$ and Nathaniel Obadia$^{3}$}

\address{$^{1}$ Universit\'e Lyon 1, Centre de Recherche Astrophysique de Lyon, 9 avenue Charles Andr\'e, F--69230 Saint--Genis--Laval, France}
\address{$^{2}$ ESA Advanced Concepts Team, ESTEC, DG-PF, Keplerlaan 1, Postbus 299, 2200 AG Noordwijk, The Netherlands}
\address{$^{3}$ \'Ecole~Normale~Sup\'erieure de Lyon, Centre de Recherche Astrophysique de Lyon, \\ 46 All\'ee d'Italie, F--69364 Lyon Cedex 07, France

\medskip
Emails: roy@obs.univ-lyon1.fr, buchert@obs.univ-lyon1.fr, sante.carloni@esa.int, nathaniel.obadia@ens-lyon.fr}

\begin{abstract}
The standard model of cosmology is based on homogeneous--isotropic solutions of Einstein's equations. These solutions are known to be gravitationally unstable to local inhomogeneous perturbations, commonly described as evolving on a background given by the same solutions. In this picture, the FLRW backgrounds are taken to describe the average over inhomogeneous perturbations for all times. We study in the present article the (in)stability of FLRW dust backgrounds within a class of averaged inhomogeneous cosmologies. We examine the phase portraits of the latter, discuss their fixed points and orbital structure and provide detailed illustrations. We show that FLRW cosmologies are unstable in some relevant cases: averaged models are driven away from them through structure formation and accelerated expansion. We find support for the proposal that the dark components of the FLRW framework may be associated to these instability sectors. Our conclusion is that FLRW cosmologies have to be considered critically for their role to serve as reliable models for the physical background.
\end{abstract}



\section{The problem of a physical background in cosmology}

The standard description of the global evolution of the universe relies on the class of homogeneous--isotropic Friedmann--Lema\^{\i}tre--Robertson--Walker (FLRW) solutions of Einstein's equations. By construction all inhomogeneities in matter and geometry, responsible for the formation of structures, average out on this given background. The conjecture is held that the FLRW background is the actual physical background.

This widely adopted framework needs to be challenged for several reasons. First, apart from the question of how to technically implement an average (a non--trivial subject for tensors), the result will depend in any case on the spatial domain over which inhomogeneities are considered: a {\it scale dependence} of the averaged variables must generically result, casting strong doubts on the scale--independent averaged values issued from the FLRW solutions (see \cite{ellisbuchert} for a summary of these thoughts). Secondly, it is actually naive to expect that a strictly homogeneous model provides the average, given the non--linearity of Einstein's equations and its implications (e.g.\ the non--commutativity of the time--evolution and averaging). This background issue may also be identified as the crucial question in the discussion of whether or not the need for dark components can be replaced by employing inhomogeneous models \cite{kolb:backgrounds}.

Looking at current cosmological structure formation models based on perturbation theories or $N$--body simulations, we find that the extra (backreaction) terms due to the averaging procedure vanish on the background of a homogeneous--isotropic solution. A closer look reveals that this property results {\it by construction} rather than by derivation, and it is a consequence of employing Newtonian or quasi--Newtonian schemes with periodic boundary conditions and a flat background geometry (see \cite{buchertehlers} for the proof). This construction is not expected to work in the framework of general relativity because of: (i) the relevance of the spatial intrinsic curvature (the second derivatives of the metric may be significant even if the metric perturbations are negligible \cite{estim}), together with (ii) the fact that inhomogeneities are coupled to the spatial curvature evolution \cite{buchert:dust}, and finally (iii) the absence of a conservation law for the averaged intrinsic curvature \cite{buchertcarfora}.

It is of course conceivable that the homogeneous solutions provide {\it in some spatial and temporal regimes} a good approximation for the evolution of the averaged distribution. A systematic approach to address this assumption is to analyze the stability of the FLRW solutions in the space of averaged cosmologies. Although FLRW solutions are known to be unstable to local perturbations, we may be able to find that they enjoy stability if subjected to perturbations on average and on some large scale. In the present paper we follow this approach, and we find that the FLRW solutions can also be globally unstable within a class of averaged models. Structure formation and accelerated expansion imply that the physical background is driven away from the FLRW background in relevant cases. Our investigation is independent of the choice of scale, and it therefore contains all the dynamical situations a domain can undergo.
 
For the class of inhomogeneous cosmologies we consider in this work, we shall employ a framework that averages the scalar parts of Einstein's equations, in which case the averaged model is unambiguously defined (reviewed in \cite{buchert:review,buchertFOCUS}) and, due to its covariant implementation, is under control \cite{buchert:fluid,li_gaugeinv,ven:cov_back,marozzi}. We are going to explore the phase space of averaged cosmologies by working within a class of exact scaling solutions, generalizing previous results \cite{morphon}\footnote{In \cite{morphon}, the scaling law was applied to the total averaged curvature, and we choose here to apply it to the averaged curvature deviation from a constant (Friedmann) curvature. Taking a null constant curvature in our work leads back to the results of \cite{morphon}. This difference of approach allows us to study in this paper the stability of all FLRW backgrounds, not fully addressed in \cite{morphon}. Note already that the phase portraits drawn in \cite{morphon} correspond to the one--dimensional locus $\Omega_k^\CD = 0$ in the plots of figure~\ref{fig:phase_portraits}.}.

The study of the phase space of dynamical systems has a long tradition. The most comprehensive study of FLRW cosmologies that include radiation, dust and a cosmological constant has been provided by Ehlers and Rindler \cite{ehlers:ps}. Other studies, including Bianchi models, were treated by Wainwright and Ellis \cite{ellisbook} who proposed the so--called dynamical system approach, which we shall use in this paper. The same technique can be followed to investigate the dynamics of minimally coupled scalar fields in the FLRW framework \cite{cop:ps}. Note that the backreaction fluid we study in this paper can also be interpreted as an effective scalar field \cite{morphon} that allows comparison of this work with our analysis. More recently, dynamical system analysis was used for modified theories of gravity (fourth order \cite{carl:rn,carl:hog,leach:2006,carloni:2007}, scalar tensor \cite{carl:stg} and Ho\v{r}ava--Lifshitz \cite{carl:hlc}) leading to a greater understanding of the cosmology of these theories. Also, results of a recent investigation of LTB models \cite{sussman:ps} are in accord with our findings\footnote{In \cite{sussman:ps}, Sussman used quasi--local variables instead of averaged ones. The relation he gave between them allows us to establish common findings.}.

In the next section, we introduce the basic framework of averaged inhomogeneous cosmologies, choosing a vanishing cosmological constant for simplicity. We consider in section~\ref{sec:scal_sol} the backreaction terms to obey scaling evolution laws, and present and analyze the corresponding autonomous dynamical system. In section~\ref{sec:disc}, we summarize and discuss the results, and we conclude in section~\ref{sec:conc}.

\section{Effective description of inhomogeneous cosmologies} \label{sec:inh_cosm}

Let us consider an {\it inhomogeneous}, irrotational and pressureless fluid (dust). The spatial average over a compact, restmass preserving domain $\CD$ of the Raychaudhuri equation, the Hamilton constraint and the dust continuity equation reads in its rest frame (see \cite{buchert:dust,buchert:fluid,buchert:review} for details)
\begin{eqnarray}
	\frac{{\ddot a}_\CD}{a_\CD} \, + \, \frac{4 \pi G}{3} \average{\varrho} \, = \, \frac{{\CQ}_\CD}{3} \, , \label{eq:eff_fried_1} \\
	\left(\frac{{\dot a}_\CD}{a_\CD}\right)^2 \, - \, \frac{8 \pi G}{3} \average{\varrho} \, + \, \frac{k_{\initial\CD}}{a^2_\CD} \, = \, -\frac{\CW_\CD \, + \, \CQ_\CD}{6} \, , \label{eq:eff_fried_2} \\
	\left\langle{\varrho}\right\rangle^{\bdot}_\CD \, + \, 3 \, \frac{{\dot a}_\CD}{a_\CD} \average{\varrho} \, = \, 0 \, , \label{eq:eff_fried_3}
\end{eqnarray}
where the angular brackets stand for the Riemannian spatial volume average, and the overdot is the partial time derivative, here identical to the covariant derivative. $a_\CD$ is the effective scale factor, $\average{\varrho}$ is the averaged energy density of the dust, $\CQ_\CD$ is the kinematical backreaction term, and $\CW_\CD$ is the averaged curvature deviation from a constant curvature, defined as
\begin{eqnarray}
	 a_\CD := \left( V_{\CD} / V_{\initial\CD} \right)^{1/3} \, , \qquad \average{\varrho} = \langle\varrho\rangle_{\initial\CD}  \, a_\CD^{-3} \, , \label{eq:def_1} \\
	{\CQ}_\CD := \frac{2}{3}\average{\left(\theta - \average{\theta}\right)^2 } - 2\average{\sigma^2} \, , \qquad \CW_\CD := \average{\CR} - 6 k_\initial\CD a_\CD^{-2} \, , \label{eq:def_2}
\end{eqnarray}
with $V_{\CD}$ being the volume of the domain, $\CR$ its three--Ricci scalar curvature and $\theta$ and $\sigma$ the expansion and shear rates of the dust. (Here and in the following, the subscript $i$ stands for the initial value.) Equation~(\ref{eq:eff_fried_2}) is an integral of equation~(\ref{eq:eff_fried_1}) under the conservation law (\ref{eq:eff_fried_3}), if and only if $\CQ_\CD$ and $\CW_\CD$ obey the integrability condition
\begin{equation}
	a_\CD^{-6} \left(\CQ_\CD \, a_\CD^6 \right)^{\bdot} + a_\CD^{-2} \left(\CW_\CD \, a_\CD^2 \right)^{\bdot} = 0 \, . \label{eq:int_cond}
\end{equation}
The set of equations (\ref{eq:eff_fried_1}-\ref{eq:eff_fried_3}) formally resembles, from a kinematical point of view, a Friedmann cosmology sourced by two fluids, but there are some fundamental differences between the two frameworks. First, the system is valid for {\it any} metric, whereas the FLRW equations only hold for maximally symmetric metrics. Second, the second fluid (the backreaction fluid) emerges {\it from the averaging process}; it is {\it not} added in our model contrary to a two--fluid FLRW cosmology. This fluid represents the difference between the dynamics of a Friedmann dust background and of a physical dust background obtained through the averaging procedure\footnote{This is an essential remark when one considers perturbations, the choice of the background being crucial.}. Third, the dynamics of the averaged geometrical variables of our model generally differs from that of the FLRW model: $\average{\CR}$ might evolve differently from the Friedmann curvature $k_\initial\CD a_\CD^{-2}$ restricted to the same domain. Fourth and last, the averaging procedure is performed over a chosen domain, and it implies a {\it scale dependence} of all the variables. In particular, one expects the backreaction fluid to exhibit rather distinct behaviors according to the scale \cite{multiscale}.

For void domains ($\varrho = 0$) the backreaction terms, encoded through geometrical invariants, correspond to inhomogeneities of the {\it spatial geometry of the vacuum}. For domains over which $\CQ_\CD = 0$, inhomogeneities could still be locally present even though the kinematical backreaction identically vanishes on $\CD$ due to an exact compensation between the expansion fluctuations and the shear\footnote{For example, this happens for zero curvature, spherically symmetric Lema\^{\i}tre--Tolman--Bondi solutions, see \cite{buchertFOCUS} in this volume.}. In the same spirit $\CW_\CD = 0$ only states that the averaged curvature deviation vanishes on $\CD$, but nothing can be inferred about the local curvature deviations within the domain. Homogeneity is only achieved, if the backreaction variables are required to vanish on all scales.

Since we aim at studying the instability of Friedmann dust backgrounds, we shall consider for our analysis the quantity $X_\CD := \CQ_\CD + \CW_\CD$, which corresponds to the {\it whole averaged departure from the Friedmann framework}. From the constraint (\ref{eq:int_cond}), a null $X_\CD$ implies vanishing $\CQ_\CD$ and $\CW_\CD$, in which case we recover a FLRW background {\it on average}. Also, the dark components can be thought of as the manifestations of the deviation of the physical background from a FLRW one: a positive $\CQ_\CD$ contributes to accelerate the expansion of the domain and plays against gravity, whereas a negative $\CQ_\CD$ participates in the deceleration of the domain's expansion and adds to gravity (see equation~(\ref{eq:eff_fried_1})). In this sense, the instability sectors, defined through $X_\CD$, represent the dark components of the concordance model.

\section{Dynamical system for the scaling solutions} \label{sec:scal_sol}

\subsection{The scaling solutions} \label{subsec:scal_sol}

We need one additional relation to solve the system (\ref{eq:eff_fried_1}-\ref{eq:eff_fried_3}), and this closure relation expresses the freedom of the choice for the local structure of the inhomogeneities, encoded on average by the dynamics of the backreaction fluid\footnote{The need for a closure relation is also present in the standard FLRW framework, and it corresponds to the choice of an equation of state for the fluid sources.}. A natural choice is given by establishing a formal analogy between the backreaction terms and a fluid, whose energy density and pressure depend on the volume of the domain\cite{morphon}:
\begin{equation}
	\hspace{-1.3cm} \CQ_\CD \propto V_\CD^\alpha \, , \qquad \CW_\CD  \propto V_\CD^\beta \qquad \Rightarrow \qquad \CQ_\CD \, = \, \CQ_{\initial\CD} a_\CD^n \, , \qquad \CW_\CD \, = \, \CW_{\initial\CD} a_\CD^p \, ,  \label{eq:scal_sol}
\end{equation}
where the scaling parameters $n$ and $p$ are real constants. Expressions~(\ref{eq:scal_sol}) together with (\ref{eq:int_cond}) result in the following constraint the dynamical system has to satisfy:
\begin{equation}
	(n + 6) \, \CQ_\initial\CD a_\CD^n \, + \, (p + 2) \, \CW_\initial\CD a_\CD^p \, = \, 0 \, . \label{eq:in_cond_scal}
\end{equation}
In what follows we shall concentrate on the class of scaling solutions $n = p$\footnote{Note that the unique solution for $n \neq p$ ($n = -6, p = -2$) is physically equivalent to the case $n = p = -6$: $\CQ_\CD$ follows the same evolution law in both situations, and so does the total averaged curvature $\average{\CR}$ which evolves proportionally to $a_\CD^{-2}$.}, for which relation~(\ref{eq:in_cond_scal}) implies
\begin{equation}
	(n + 2) \CW_\CD = - (n + 6) \CQ_\CD \quad \Rightarrow \quad (n + 2) X_\CD = - 4 \CQ_\CD \, . \label{eq:x_fluid}
\end{equation}

\subsection{The autonomous system}

Let us introduce the Hubble functional $H_\CD := \dot{a}_\CD / a_\CD$ and the volume deceleration functional $q_\CD := - 1 - \dot{H}_\CD/H^2_\CD$. We define {\it domain--dependent} dimensionless cosmological characteristics as
\begin{equation}
	\Omega_m^\CD := \frac{8 \pi G}{3 H_\CD^2} \langle\varrho\rangle_{\cal D}  \, , \qquad
	\Omega_k^\CD := - \frac{k_{\initial\CD}}{a^2_\CD \, H_\CD^2} \, ,\qquad
	\Omega_X^\CD := - \frac{X_\CD}{6 H_\CD^2 } \, , \label{eq:omega}
\end{equation}
which by construction add up to 1 according to the Hamilton constraint (\ref{eq:eff_fried_2}). Upon using the expressions~(\ref{eq:x_fluid},\ref{eq:omega}) we replace the system (\ref{eq:eff_fried_1}-\ref{eq:eff_fried_3},\ref{eq:int_cond}) by
\begin{eqnarray}
	\Omega_m^\CD - (n + 2) \Omega_X^\CD = 2 q_\CD \, , \label{eq:dyn_sys1} \\
	\Omega_m^\CD + \Omega_k^\CD + \Omega_X^\CD = 1 \, , \label{eq:dyn_sys2} \\
	{\Omega_m^\CD}^{'} = \Omega_m^\CD \left(\Omega_m^\CD \, - \, (n + 2) \, \Omega_X^\CD \, - \, 1 \right) \, , \label{eq:dyn_sys_a} \\
	{\Omega_X^\CD}^{'} = \Omega_X^\CD \left(\Omega_m^\CD \, - \, (n + 2) \, \Omega_X^\CD \, + \, n \, + \, 2 \right) \, , \label{eq:dyn_sys_b}
\end{eqnarray}
where the prime denotes the derivative with respect to the evolution parameter \hbox{$N_\CD := \ln{a_\CD}$}. The two last equations are easily derived noting that $(1/H_\CD^2)' = 2(q_\CD + 1)/H_\CD^2$. The autonomous system (\ref{eq:dyn_sys_a},\ref{eq:dyn_sys_b}) determines the orbit of a cosmological state, defined by the quartet $(\Omega_m^\CD,\Omega_X^\CD,\Omega_k^\CD,n)$, in the corresponding phase space, and its fixed points are provided in table~\ref{tab:fixed_points}. The phase space $(\Omega_m^\CD,\Omega_X^\CD)$ presents three invariant lines under the phase flow: $\Omega_m^\CD = 0$, $\Omega_X^\CD = 0$ and $\Omega_k^\CD = 1 -  \Omega_m^\CD - \Omega_X^\CD = 0$.
\begin{table}[!htb]
\caption{Fixed points of the dynamical system (\ref{eq:dyn_sys_a},\ref{eq:dyn_sys_b}). The sign of their eigenvalues determines their stability \cite{arrow}.}
	\center{
		{\footnotesize
		\begin{tabular}{cccc}
			\hline
			& {\bf Coordinates} & \hspace{-3mm} \multirow{2}{*}{\bf Scale factor} & \hspace{-2mm} {\bf Stability} \\ 
			& $(\Omega_m^\CD, \Omega_X^\CD)$ & & \{eigenvalues\}\\
			\hline \\
			point {$\mathcal{A}$} & {$(0,0)$} & 
			\hspace{-3mm} $a_\CD(t) = H_\initial\CD \, (t-t_i) \, + \, 1$ &
			\hspace{-2mm} $\begin{array}{c}
				n < -2, \; \mathrm{attractor} \\
				n > -2, \; \mathrm{saddle} \\
				\{-1,n+2\}
			\end{array}$
			\\
			\\
			point {$\mathcal{B}$} & {$(1,0)$} & 
			\hspace{-3mm} $a_\CD(t) = \left( \frac{3}{2} H_\initial\CD \, (t-t_i) \, + \, 1 \right)^{\textstyle \frac{2}{3}}$ &
			\hspace{-2mm} $\begin{array}{c}
				n < -3, \; \mathrm{saddle} \\
				n > -3, \; \mathrm{repeller} \\
				\{1,n+3\}
			\end{array}$
			\\
			\\
			point {$\mathcal{C}$} & {$(0,1)$} &
			\hspace{-3mm} $\begin{array}{c}
			a_\CD(t) = \left( - \frac{n}{2} H_\initial\CD \, (t-t_i) \, + \, 1 \right)^{\textstyle - \frac{2}{n}} \quad (n \neq 0) \vspace{2mm} \\
			a_\CD(t) = \exp({H_\initial\CD \, (t-t_i)}) \quad (n = 0) 
			\end{array}$ &
			\hspace{-2mm} $\begin{array}{c}
				n < -3, \; \mathrm{repeller} \\
			\!	-3 < n < -2, \; \mathrm{saddle} \\
				n > -2, \; \mathrm{attractor} \\
				\{- n - 3,- n - 2\}
			\end{array}$
			\\
			\\
			\hline \\
			line $\mathcal{L}_1$ & \;\; $\Omega_m^\CD + \Omega_X^\CD = 1$ &
			$a_\CD(t) = \left( \frac{3}{2} H_\initial\CD \, (t-t_i) \, + \, 1 \right)^{2/3}$ &
			$(n = -3)$
			\\
			\\
			line $\mathcal{L}_2$ & \;\; $\Omega_m^\CD = 0$ &
			$a_\CD(t) = H_\initial\CD \, (t-t_i) \, + \, 1$ &
			$(n = -2)$
			\\
			\\
			\hline
		\end{tabular}
		}
	\label{tab:fixed_points}
	}
\end{table}

There are five situations for which the stability of the fixed points does not change: $n <-3$, $n = -3$, $-3 < n < -2$, $n = -2$ and $n > -2$. The fixed point $\mathcal{A}$ shares the same properties as a Milne domain (i.e.\ a void domain with FLRW kinematics and FLRW curvature on average); the fixed point $\mathcal{B}$ corresponds to an Einstein--de Sitter dust domain (i.e.\ FLRW kinematics and zero FLRW curvature), and the fixed point $\mathcal{C}$ corresponds to domains filled only with the backreaction fluid. The stability of $\mathcal{A}$ and $\mathcal{B}$ can change according to the value of the scaling parameter, but they always depict the same cosmological models. In return, $\mathcal{C}$ always describes different cosmological models which might also have a different stability. For the bifurcation value $n = -3$, there is a transfer of stability between $\mathcal{B}$ and $\mathcal{C}$ which both belong to the line of fixed points $\mathcal{L}_1$, while at the bifurcation value $n = -2$, there is a transfer of stability between $\mathcal{A}$ and $\mathcal{B}$, both belonging to the line $\mathcal{L}_2$.

Finally, note that the dynamical system (\ref{eq:dyn_sys_a},\ref{eq:dyn_sys_b}) is written according to the evolution parameter $N_\CD$ and {\it not according to a time parameter}. Consequently, the phase portraits drawn from this system do not suffice to deduce the {\it time evolution} of a cosmological state, since we also need information about the time evolution of $N_\CD$ or, equivalently, $a_\CD$. We shall therefore also indicate for all phase portraits the behavior of the scale factor.

\subsection{Phase portraits}

\begin{figure}[!ht]
	\center{\includegraphics[scale=0.7]{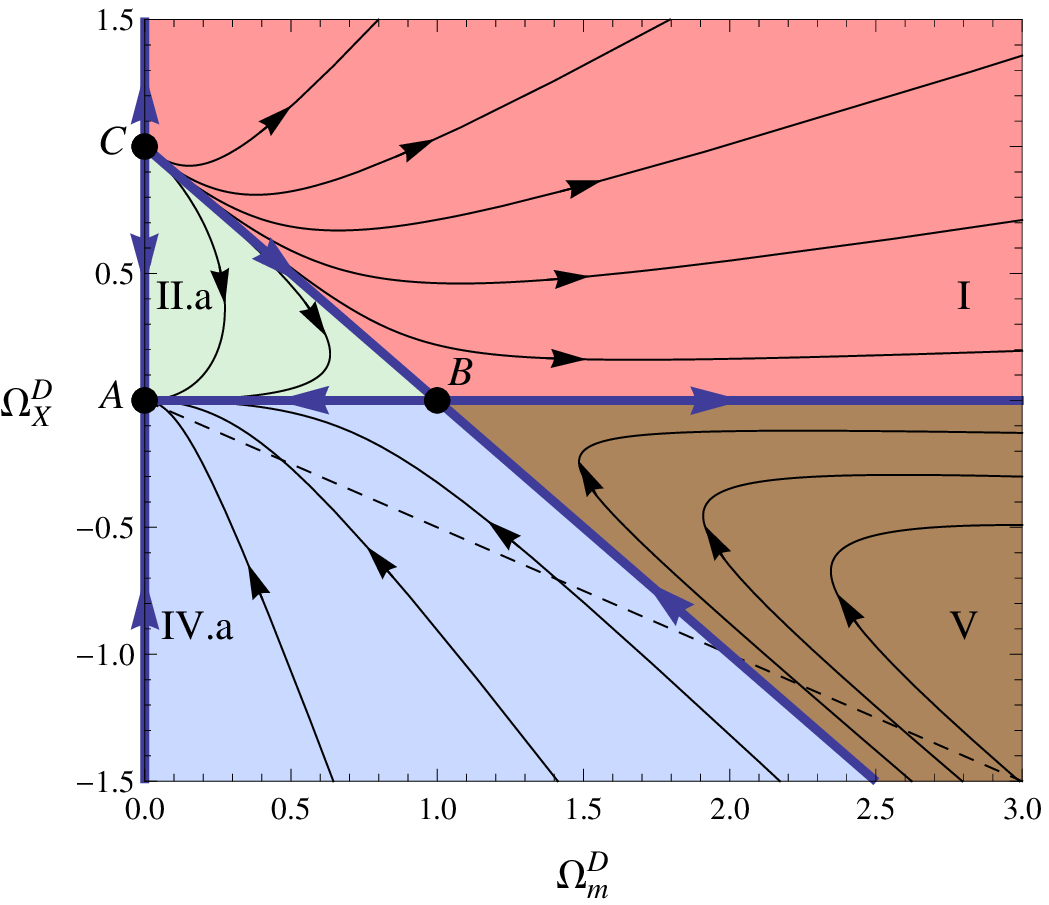} \hspace{5mm}
	\includegraphics[scale=0.7]{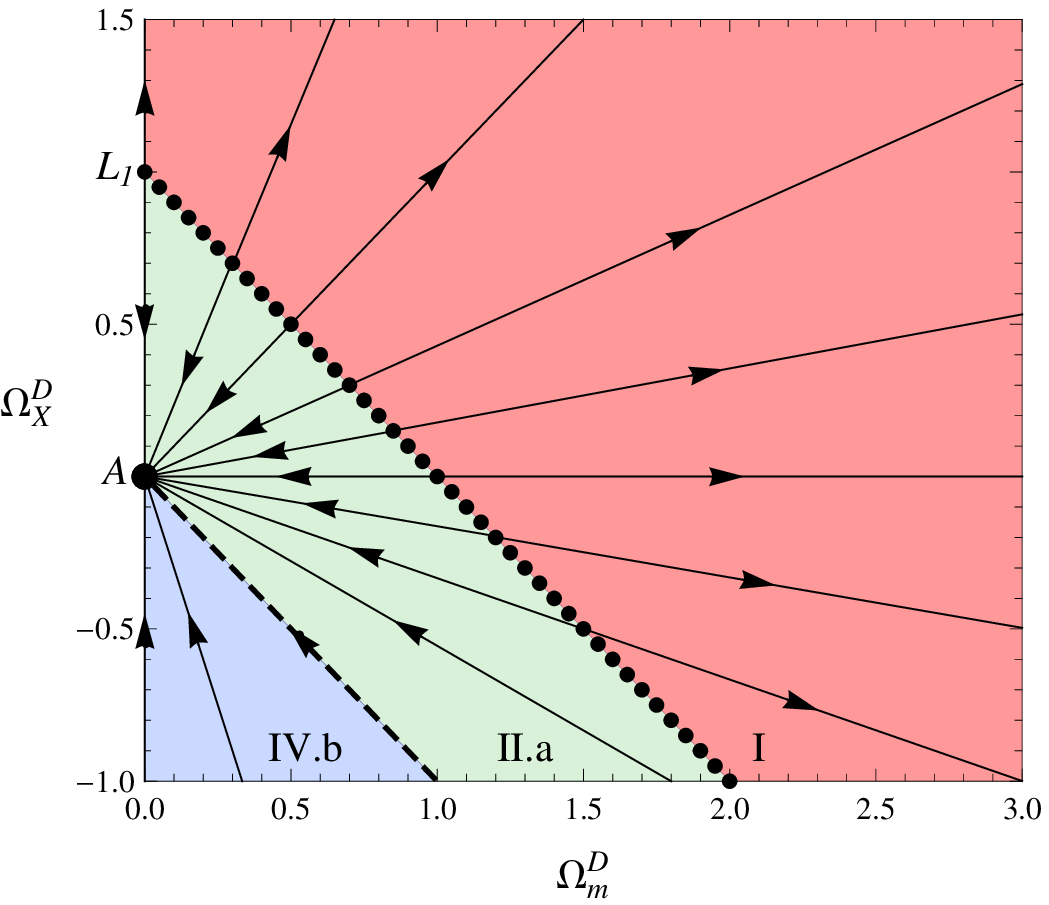} \\
	\includegraphics[scale=0.7]{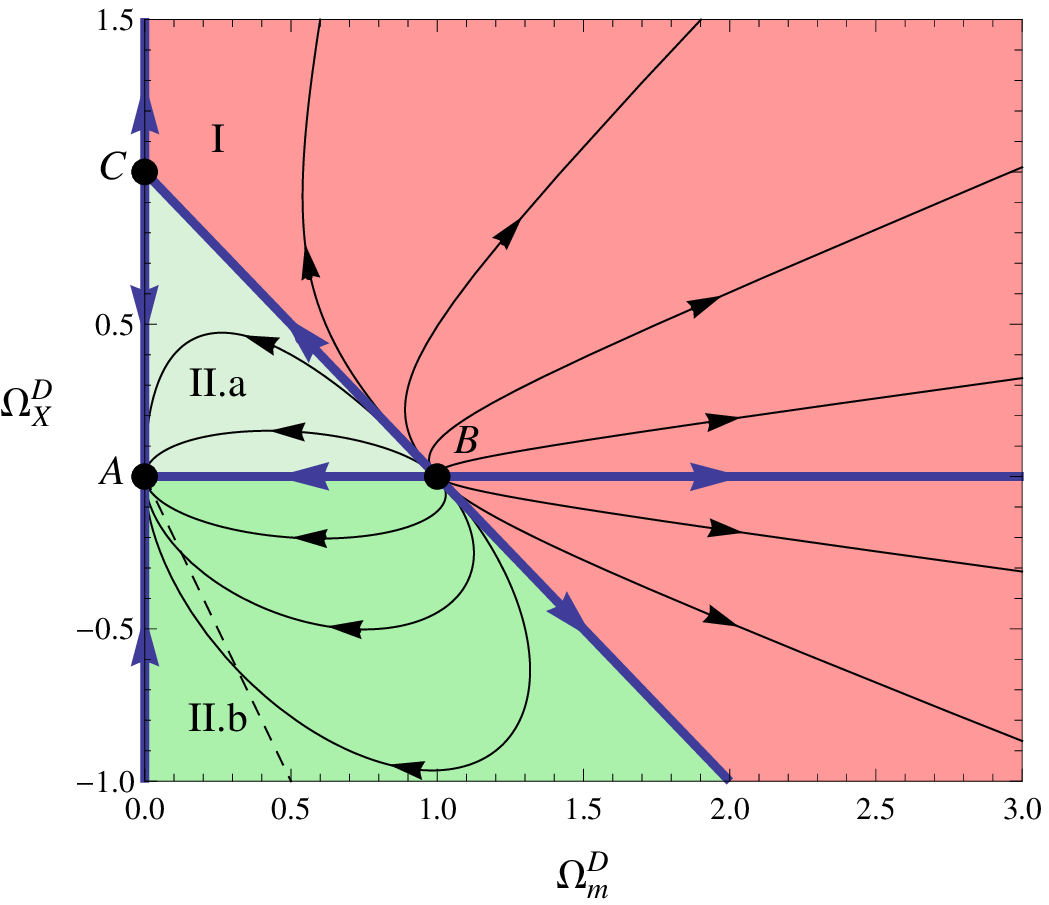} \hspace{5mm}
	\includegraphics[scale=0.7]{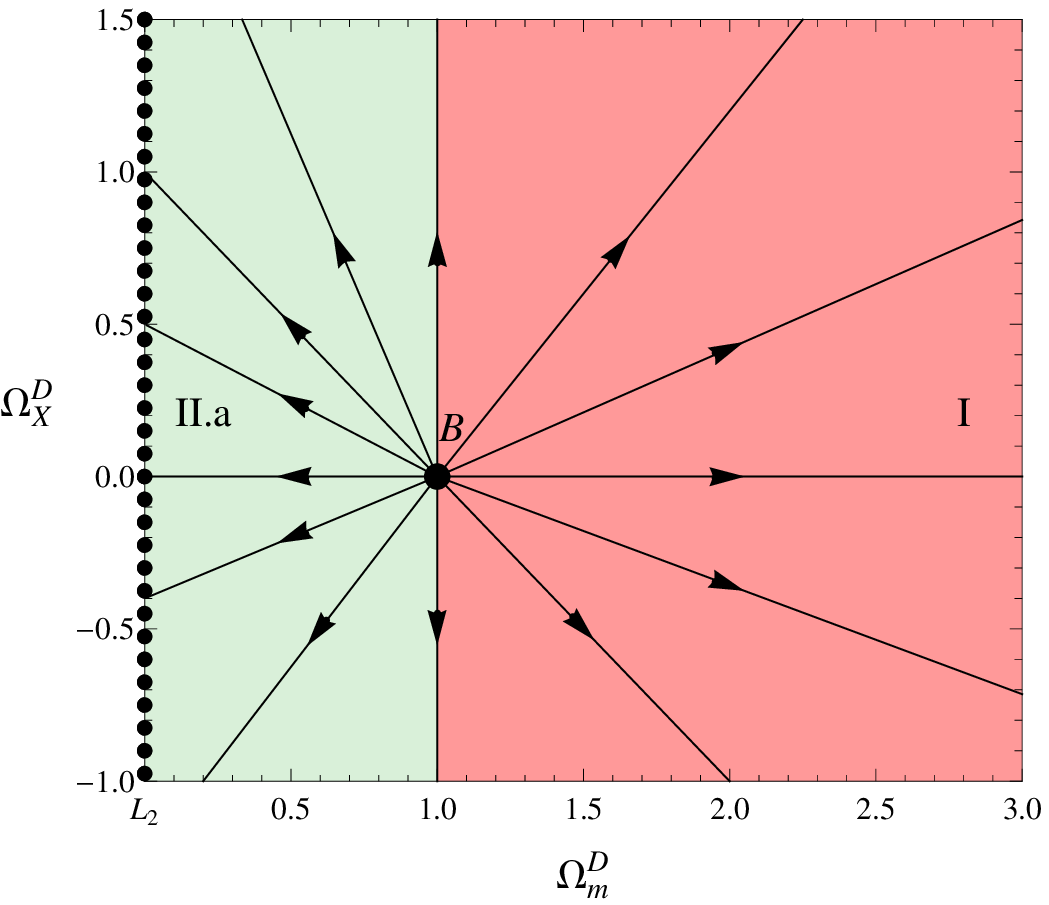} \\
	\includegraphics[scale=0.7]{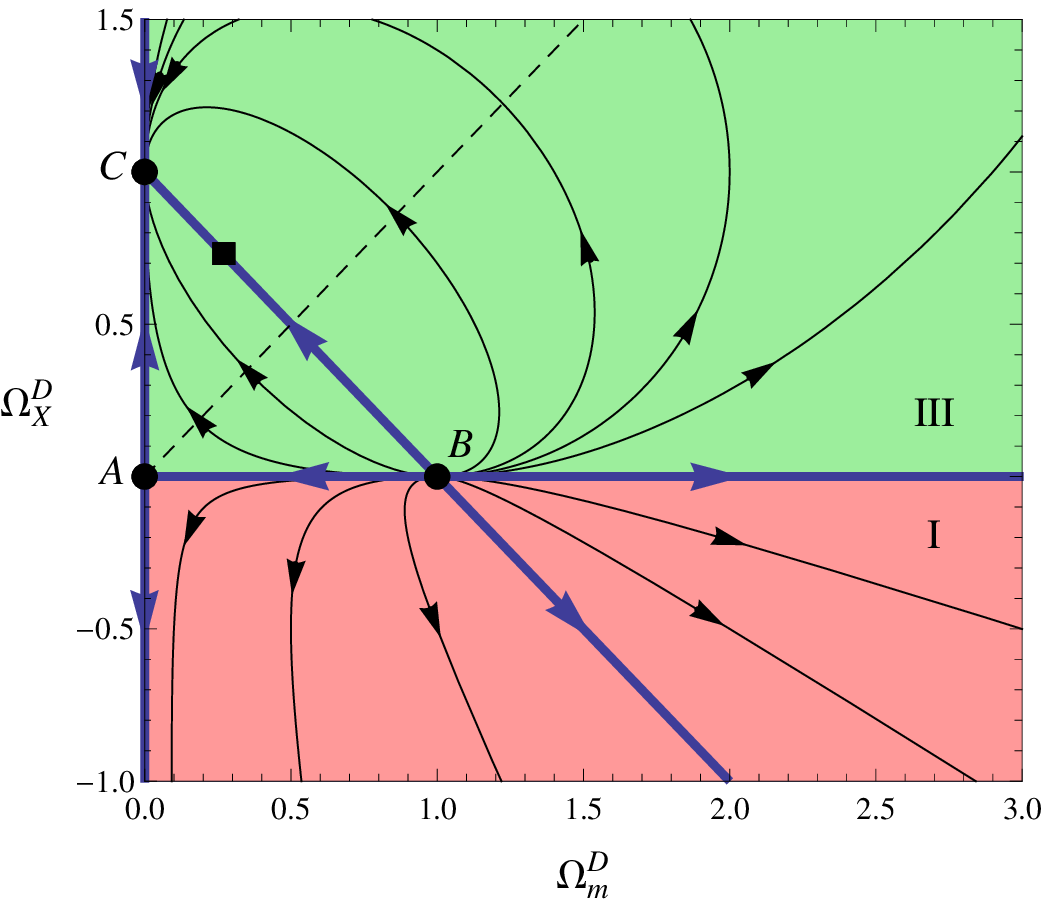}
	}
	\vspace{-3mm}
	\caption{Phase space $(\Omega_m^\CD,\Omega_X^\CD)$ of the dynamical system (\ref{eq:dyn_sys_a},\ref{eq:dyn_sys_b}). From left to right and top to bottom: $n < -3$ ($n = -4$ for the illustration), $n = -3$, $-3 < n < -2$ ($n = -2.5$), $n = -2$ and $n > -2$ ($n = -1$). The thick straight lines (dark blue) are the invariant lines $\Omega_m^\CD = 0$, $\Omega_X^\CD = 0$ and $\Omega_k^\CD = 1 - \Omega_m^\CD - \Omega_X^\CD = 0$. Every line parallel to the latter, the inclined line, corresponds to a constant $\Omega_k^\CD$, with values increasing downward. The dashed line shows a vanishing $q_\CD$; it is an orbit only for $n = -3$, and it coincides with $\Omega_m^\CD = 0$ for $n = -2$. A domain is accelerated below this line and decelerated above when $n < -2$, and this correspondence is inverted for $n > -2$. The arrows show an increasing $N_\CD$, the dots are the fixed points of table~\ref{tab:fixed_points}, and the square in the last plot stands for the observational values (see section~\ref{subsec:dark}).}
	\label{fig:phase_portraits}
\end{figure}

\begin{figure}[!ht]
	\center{\includegraphics[scale=0.6]{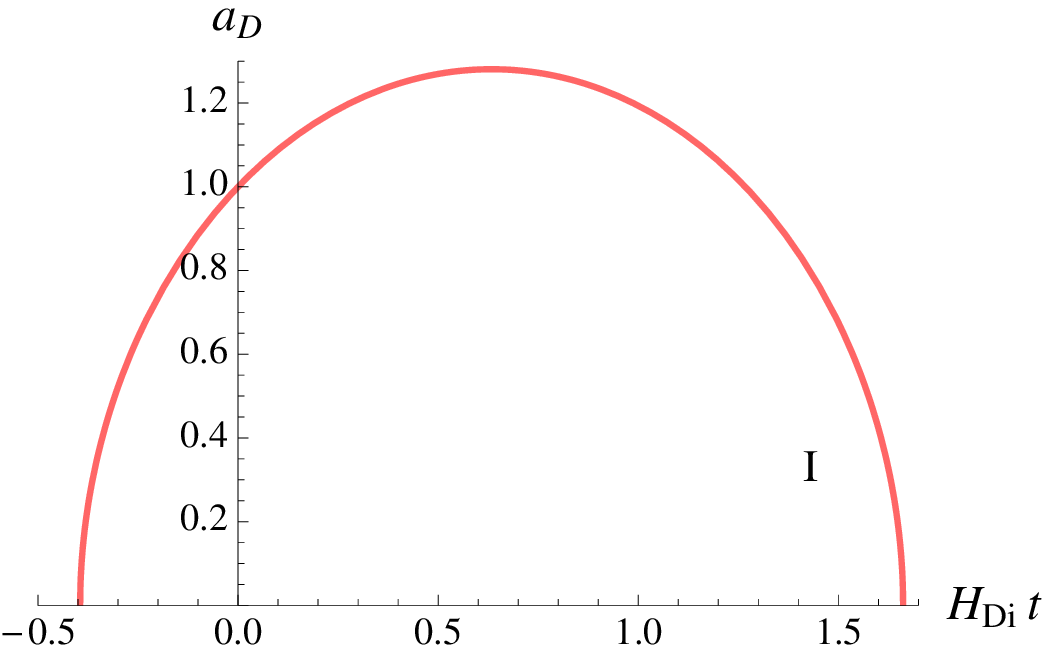} \hspace{7mm}
	\includegraphics[scale=0.6]{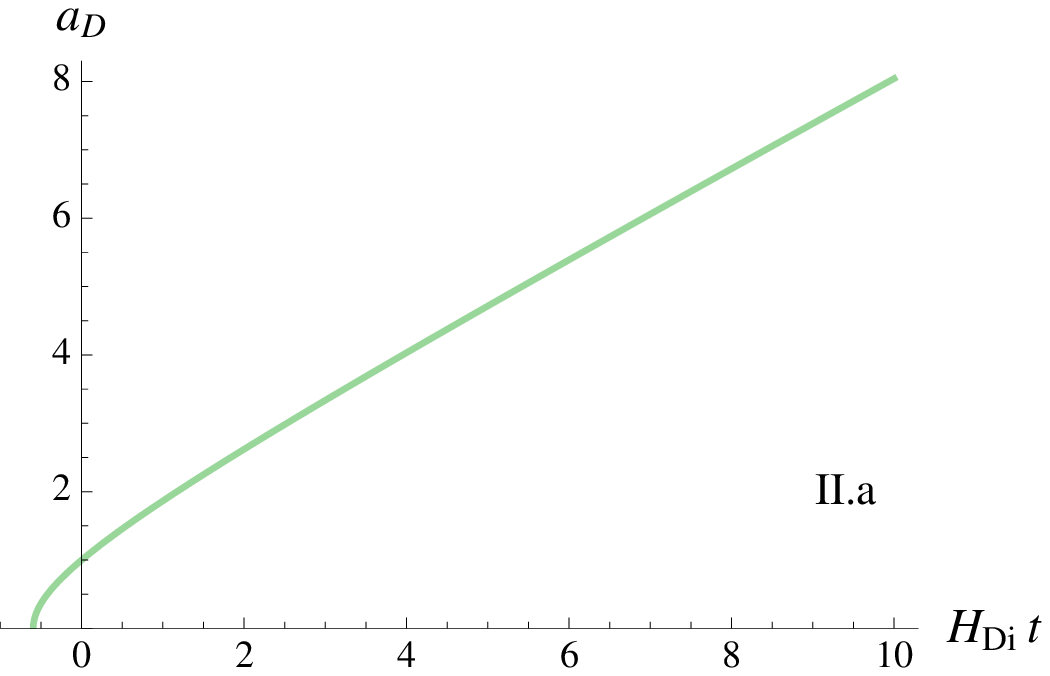} \\ \vspace{3mm}
	\includegraphics[scale=0.6]{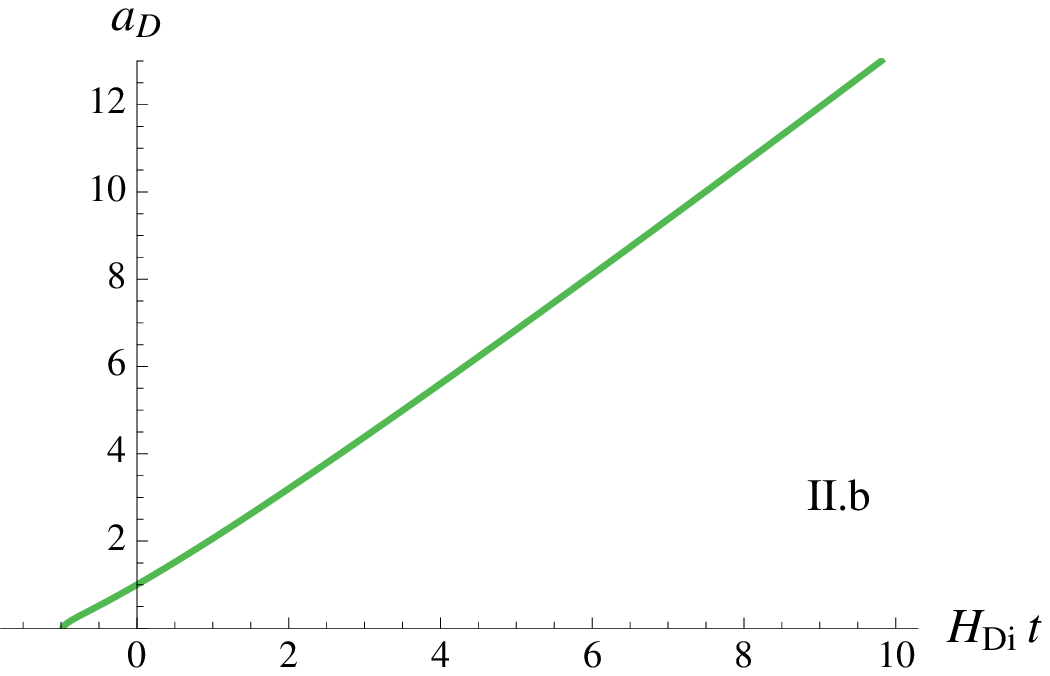} \hspace{7mm}
	\includegraphics[scale=0.6]{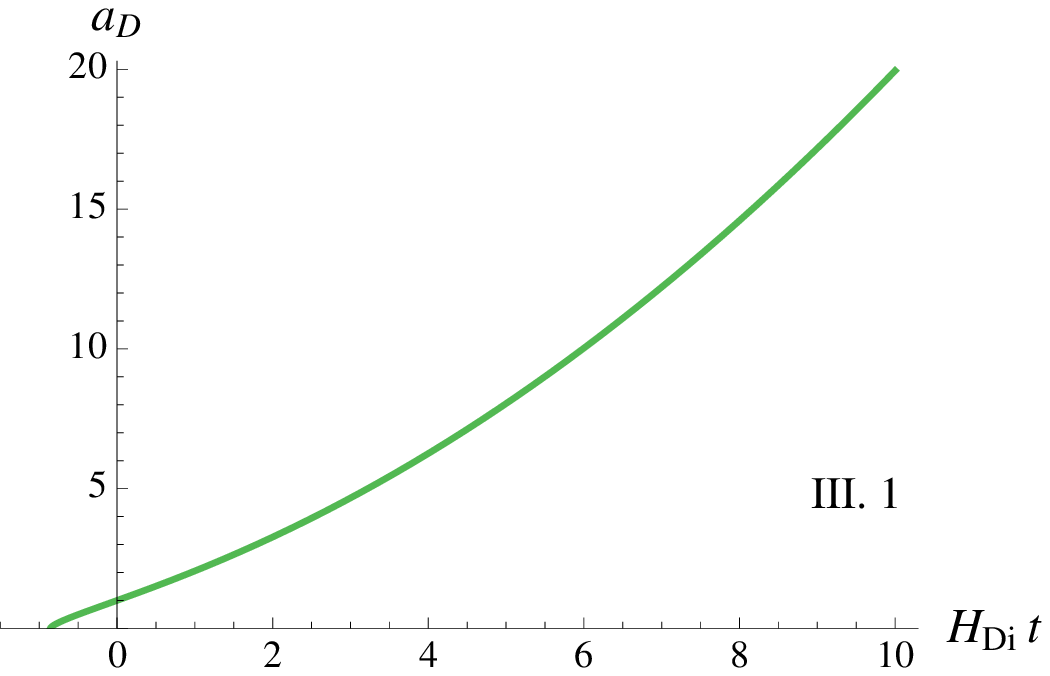} \\ \vspace{3mm}
	\includegraphics[scale=0.6]{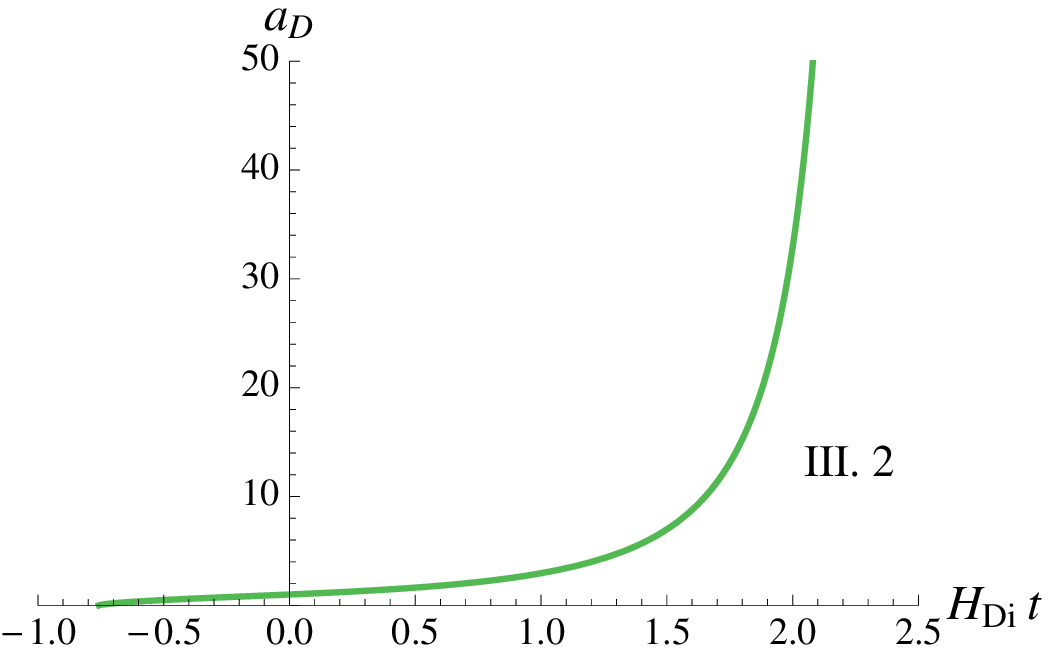} \hspace{7mm}
	\includegraphics[scale=0.6]{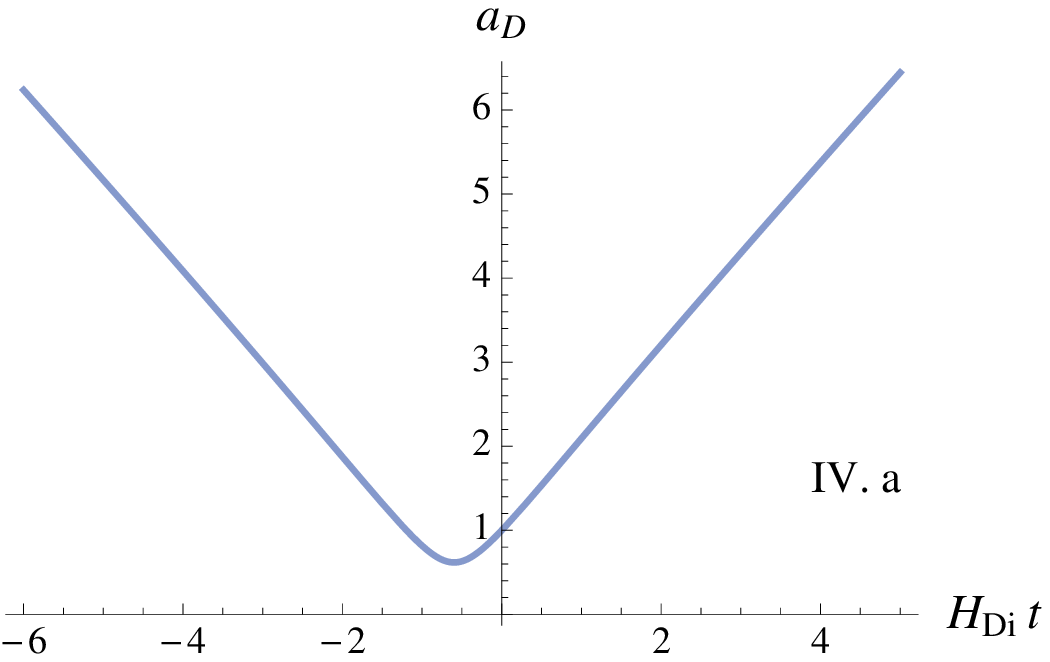} \\ \vspace{3mm}
	\includegraphics[scale=0.6]{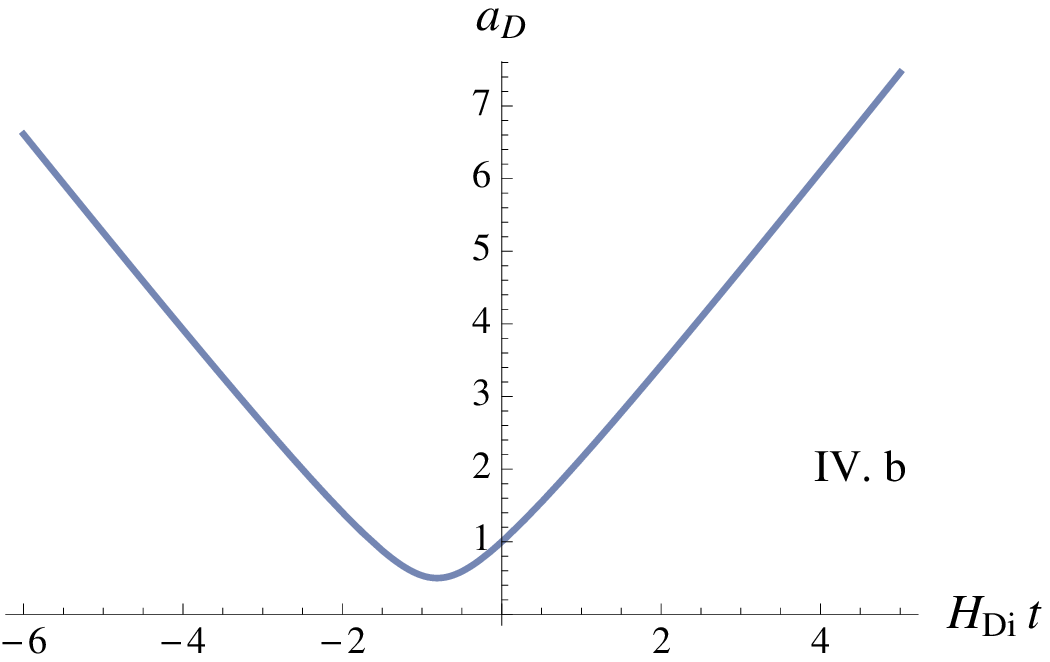} \hspace{7mm}
	\includegraphics[scale=0.6]{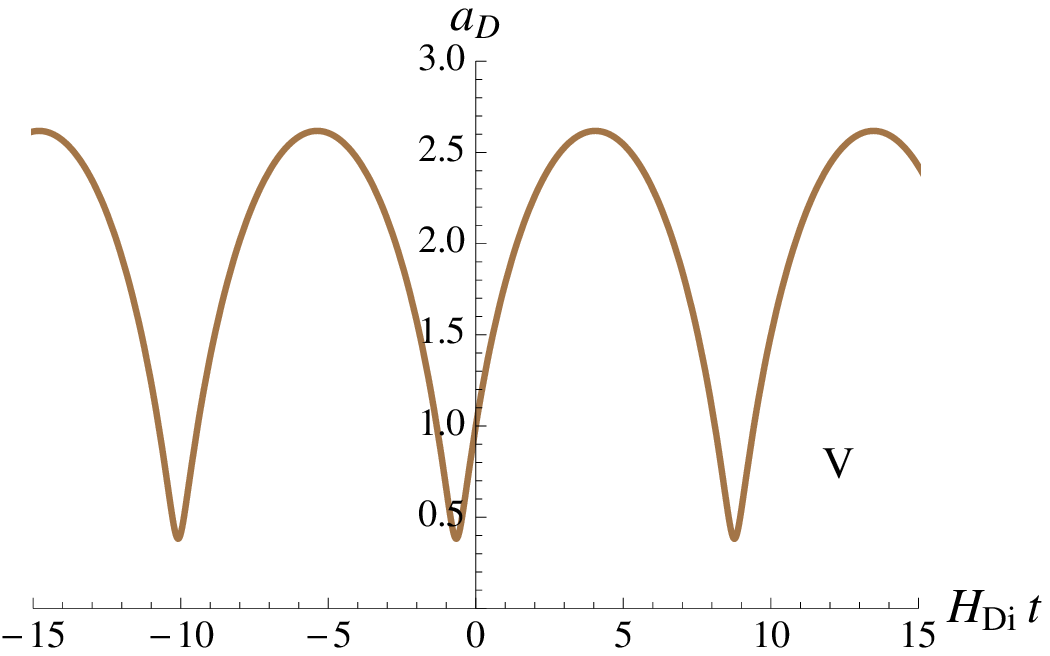}
	}
	\caption{Evolution of the scale factor of a domain with respect to $H_\initial\CD t$. The numbers (colors) refer to the different regions of the phase spaces in figure~\ref{fig:phase_portraits}. All the states of a region have a scale factor which evolves in the same way. We have chosen a positive $H_\initial\CD$ to deal with initially expanding domains and a null $t_i$. Region I (red) shows big--bang and big--crunch domains, regions II show big--bang and future expanding (with an asymptotically null acceleration) domains, region III.1 (darkest green) describes big--bang and future expanding (accelerating) domains for $-2 < n \leq 0$, region III.2 (darkest green) describes big--bang and big--rip domains for $n > 0$, regions IV (blue) show past and future expanding domains, and region V (brown) shows oscillating domains. We distinguish two subcases for the regions II and IV: for the former, region II.a is never accelerated, while region II.b is accelerated from a finite time onward; for the latter, region IV.a is accelerated during a finite period of time, while region IV.b is always accelerated.}
	\label{fig:scale_fact}
\end{figure}

The phase portraits of the five cases are shown in figure~\ref{fig:phase_portraits}, and the numbers (background colors) of their regions correspond to the evolutions of $a_\CD$ given in figure~\ref{fig:scale_fact}. To deduce the time evolution of a state, one shall refer to these two sets of plots. For the sake of clarity, we provide hereafter a detailed explanation of the correspondence between them in the situation $n < -3$ (first plot of figure~\ref{fig:phase_portraits}).

A state $(\Omega_m^\CD,\Omega_X^\CD)$ belonging to the region I (red) evolves from the fixed point $\mathcal{C}$ when its scale factor is zero (plot I (red) of figure~\ref{fig:scale_fact}), through its initial conditions, to infinity when its scale factor attains its maximum. This divergence is due to the singularity of the cosmological parameters for~$H_\CD\,$=$\,$0~and not due to a physical boundless blow of the domain. Having reached its maximal extension, the domain shrinks and $(\Omega_m^\CD,\Omega_X^\CD)$ comes back from infinity and tends to $\mathcal{C}$, while $a_\CD$ decreases from its maximum to zero. Hence, the fixed point $\mathcal{C}$ is both a past repeller and a future attractor.

In the region II.a (green), the states evolve from $\mathcal{C}$, when $a_\CD$ is zero, toward $\mathcal{A}$, when $a_\CD$ tends to infinity (plot II.a (green) of figure~\ref{fig:scale_fact}), without any accelerated period. In this region, $\mathcal{C}$ is a past repeller and $\mathcal{A}$ is a future attractor.

A state in IV (blue) leaves the fixed point $\mathcal{A}$, while the scale factor decreases from infinity to its minimum in the past, and reaches infinity. It then goes back to $\mathcal{A}$ when the scale factor grows to infinity (plot IV.a (blue) of figure~\ref{fig:scale_fact}). The state undergoes an accelerated period when operating its bouncing back from infinity. Here, $\mathcal{A}$ is a past repeller and a future attractor.

Finally, in the region V (brown), $(\Omega_m^\CD,\Omega_X^\CD)$ runs from infinity at the minimum of $a_\CD$ to infinity at its maximum, and it oscillates (plot V (brown) of figure~\ref{fig:scale_fact}). The minimum of $a_\CD$ is reached when the orbit runs below the dashed line since it corresponds to a positive acceleration.

All the orbits that go to infinity only describe part of the evolution, i.e.\ up to a maximum or minimum of the scale factor. The rest of the evolution is described not by reversing the direction of time, but by reversing the monotonicity of $a_\CD$. In other words, the state travels back on the orbit of the phase portraits.

The other generic orbits expounded in figure~\ref{fig:phase_portraits} are the following (see also figure~\ref{fig:scale_fact}): a state in IV.b (blue) follows the same type of evolution as a state in IV.a except that it always undergoes an acceleration; the evolution of a state in II.b (green) differs from that of II.a by the fact that it experiences an acceleration from a finite time; and a state in III (darkest green) evolves from $\mathcal{B}$, when $a_\CD$ is zero, toward $\mathcal{C}$, when $a_\CD$ tends to infinity. In this region, $\mathcal{B}$ is a past repeller and $\mathcal{C}$ is a future attractor.

Finally, we present the evolution of the scale factor for domains lying on an invariant line. If a state belongs to the segment between two fixed points, it experiences a big--bang and a future expanding phase. If it belongs to a semiline, either the domain undergoes a big--bang and big--crunch evolution when the arrow on the invariant line leaves the fixed point, or the domain undergoes a past and future expanding phase when the arrow is orientated toward the fixed point.

\FloatBarrier

\section{Summary and discussion} \label{sec:disc}

First and foremost, we clarify the definitions of the different types of domains at stake. A domain $\CD$ is {\it Friedmann}, i.e.\ locally homogeneous and isotropic, if and only if for all subdomains $\CE \subseteq \CD$ one has $\Omega_\CQ^{\CE} =\Omega_\CW^{\CE} = 0$ or, equivalently, $\Omega_X^{\CE} = 0$. We have focused our analysis on one single domain $\CD$ without any multiscale considerations; so it is fair to assert that the Friedmann realization is not achieved in our approach since no additional suppositions are made. A domain with a null $\Omega_X^\CD$ and with nothing more inferred about the $\Omega_X^{\CE}$ of the subdomains, is most probably non--Friedmann: as noted in section~\ref{sec:inh_cosm}, the cancellation of the backreaction fluid can be the result of an exact compensation between the expansion variance and the shear, and inhomogeneities and anisotropies may still be locally present. Nevertheless, such domains are destined to follow {\it on average} a Friedmann evolution since their inhomogeneities do not contribute {\it on average} to the kinematics. We shall call these domains {\it Friedmann--like} domains. Therefore, we shall deal with {\it Milne--like} and {\it Einstein--de Sitter--like} domains, instead of locally Milne and Einstein--de Sitter domains. This complies with the actual physical situation: Friedmann models as dynamical models for the average evolution are only realized for strictly homogeneous distributions of matter and geometry, but there exist situations---called {\it Friedmann--like}---in which Friedmann models do provide the evolution on average, i.e. in which they describe the {\it physical background}.

Another terminology we employ makes use of equation~(\ref{eq:x_fluid}): for $n<-2$, $\Omega_X^\CD > 0$ corresponds to shear--dominated domains, and $\Omega_X^\CD < 0$ labels expansion variance--dominated domains. This correspondence is inverted for $n > -2$.


\subsection{Summary of results}

Before delving into details, simply recall that the orbits strongly depend on the leading term in the cosmic trio $\Omega_m^\CD \propto  a_\CD^{-3} H_\CD^{-2}$, $\Omega_k^\CD \propto  a_\CD^{-2} H_\CD^{-2}$ and $\Omega_X^\CD \propto  a_\CD^{n} H_\CD^{-2}$.  When $n < -2$, the constant curvature $k_\initial\CD$, useful to compare the effective background with the Friedmann--like background, determines the dynamics of the domain, while in the remaining situations the backreaction fluid is the relevant quantity.

For $n < -3$ and for a negative constant--curvature $k_\initial\CD$, shear--dominated and expansion variance--dominated domains are attracted by a Milne--like state. In the past, the former originates from a state with $\Omega_X^\CD \simeq 1$, whereas the latter comes from a Milne--like state. Expansion variance--dominated domains are therefore asymptotically, in the past and in the future, dominated by a negative constant curvature. The bounce of these domains occurs when the contribution of the backreaction fluid is no longer negligible. For a positive $k_\initial\CD$, expansion variance--dominated domains are oscillating (this is the only situation where it occurs): when the volume of the domain is sufficiently small, the backreaction fluid---mimicking a dark energy behavior---becomes preponderant over the other components, thus avoiding the collapse. In contrast, shear--dominated domains---mimicking a dark matter behavior---evolve toward a domain filled only with $\Omega_X^\CD$ and collapse in a finite time.


At $n = -3$, all domains with a negative $k_\initial\CD$ approach a Milne--like state, and all domains with a positive $k_\initial\CD$ eventually collapse. The attractor state of these latter depends on the initial conditions: though it always exhibits $\Omega_k^\CD = 0$, all the values of $\Omega_m^\CD$ and $\Omega_X^\CD$ are attainable provided the Hamilton constraint is satisfied. For the particular state $\Omega_k^\initial\CD = 1$ (dashed line of the second plot of figure~\ref{fig:scale_fact}), the domain experiences a stationary expansion ($\dot{a}_\CD=\mathrm{const}$). $\Omega_k^\CD$ always being equal to $1$ in this situation, the dust and backreaction fluid energies exactly compensate each other along the whole evolution. Furthermore, we here confirm that this stationary state is {\it stable}: any perturbation always leads back to a stationary state (as was conjectured in \cite{buchert:static}, see also \cite{barrowetal:static,unruh:static} for comparisons).

For $-3 < n < -2$, shear--dominated and expansion variance--dominated domains have the same qualitative asymptotic properties. For a negative $k_\initial\CD$, they originate from an Einstein--de Sitter--like dust state and are attracted by a Milne--like state, whereas for a positive constant curvature they emanate from and converge to an Einstein--de Sitter--like state and eventually collapse. In these last scenarios, even if the backreaction fluid mimics a dark energy behavior over the domain, its intensity is not sufficient enough to counterbalance the collapse contrary to the case $n < -3$.


In all the previous situations (except for the region V (brown)), a positive constant--curvature domain is bound to collapse, and a negative constant--curvature domain always converges toward a Milne--like state. Therefore, a small perturbation around a null $k_\initial\CD$ seals a totally different fate according to its sign.

For $n = -2$, domains with $\Omega_m^\initial\CD < 1$ are attracted by a Milne--like state, and domains with $\Omega_m^\initial\CD > 1$ converge to an Einstein--de Sitter--like state and collapse within a finite time. The constraint~(\ref{eq:int_cond}) here implies $\CQ_\CD = 0$ and $\CW_\CD \propto a_\CD^{-2}$, which means that there is no kinematical backreaction, but also no curvature deviation from a constant curvature {\it stricto sensu}. All states follow the kinematics of a Friedmann--like state on average, and the whole drawn phase portrait is physically equivalent to the line $\Omega_X^\CD = 0$.

Finally, for $n > -2$, shear--dominated domains emerge from and are attracted by an Einstein--de Sitter--like dust state, and they eventually collapse. Expansion variance--dominated domains also originate from an Einstein--de Sitter--like state, but they are attracted by a state filled only with the backreaction fluid, and they experience an accelerating expansion from a finite time onward. As an example for the latter: void domains that are nearly Friedmann on average ($\Omega_m^{\CD} = 0$, $\Omega_X^\CD \gtrsim 0$) are unstable and driven away toward this expanding state; this instability is the origin of the possible onset of an inflationary scenario created from initial curvature inhomogeneities of the Einstein vacuum \cite{infl:bu_ob}.

\subsection{Discussion: instability sectors} \label{subsec:inst}

A state leaves the class of FLRW {\it backgrounds} and enters into what we define as an {\it instability sector}\footnote{From figure~\ref{fig:phase_portraits}, it is obvious that no Friedmann--like {\it state} is stable (except the point ${\mathcal A}$), i.e. in those unstable cases no perturbation slightly above the $\Omega^X_\CD=0$ line relaxes to its original point of the phase space diagram. Therefore, we address the more general issue of stability of the {\it class of backgrounds} with Friedmann--like behavior.} as soon as the backreaction fluid contributes to the kinematics of the domain. The existence of a non--vanishing averaged contribution of inhomogeneities invokes deviations of a cosmic state from its Friedmann--like characteristics. Given the generic behavior of any domain, we now address the issue of the stability of the FLRW backgrounds: does a Friedmann--like state, lying on the line $\Omega_X^\CD = 0$, converge {\it on average} to the same or another Friedmann--like state when subjected to perturbations?

A convergence toward any Friedmann--like state or, equivalently, toward the class of FLRW backgrounds, happens when the two following conditions are met: first, $X_\CD\to 0$, which entails that inhomogeneities are negligible {\it on average}; second, $\Omega_X^\CD \to 0$, which states that the energy contribution of the backreaction fluid becomes subdominant {\it on average} compared to the other components. Note that for the particular case $n = -2$, as explained above, the averaged contribution of inhomogeneities vanishes, and all the states are Friedmann--like. This situation does not admit any global inhomogeneity effect, and we shall therefore disregard it in our discussion.

Based on the previous analyses, we conclude that FLRW backgrounds are stable for: (i) $n < -2$ and $k_\initial\CD < 0$, the Milne--like state is an attractor for any inhomogeneous deviations, and (ii) $n < -3$ and $k_\initial\CD = 0$, the Einstein--de Sitter--like dust state is eventually reached. In these situations the FLRW backgrounds are stable, in the situation (i) the background can only be attained asymptotically. 

In the other situations, the perturbations of any Friedmann--like state lead either to $\Omega_X^\CD \to 1$ or $X_\CD \to \infty$ or both, namely to a non Friedmann--like state. The FLRW backgrounds are therefore globally unstable for: (iii) $n < -2$ and $k_\initial\CD > 0$, (iv) $-3 \leq n < -2$ and $k_\initial\CD = 0$, and (v) $n > -2$. In these situations they obviously constitute an incorrect approximation to the physical background.

\subsection{Discussion: interpretation as dark sectors} \label{subsec:dark}

The different cosmological observations interpreted in the context of FLRW cosmologies constrain the instability sectors to which the dark components could be associated. For instance, we may put a conservative upper bound on the present--day cosmic state, $\Omega_{m, \mathrm{today}}^\CD < 1$, i.e. any physical orbit should respect this bound, if we consider small--scale (galaxy clusters and voids) or large--scale (CMB and high--redshift supernovae) cosmological domains. For instance, this restriction forbids some of the orbits for these gravitational systems: they cannot belong to the region V (brown) of figure~\ref{fig:phase_portraits}, and so we cannot have oscillating domains.

The {\it dark matter} component can be associated with domains that are dominated by shear fluctuations, and which eventually collapse. The instability sectors of the regions I (red) in the first plot of figure~\ref{fig:phase_portraits}, in the second and third plots (with $\Omega_X^\CD > 0$), and in the fifth plot (with $\Omega_X^\CD < 0$) may present an effective dark matter behavior for the backreaction fluid. Concerning the {\it dark energy} component, it is effectively mimicked in the instability sectors of the region IV.b (blue) in the second plot, II.b (green) in the third plot, and III (darkest green) in the last plot. In each of these cases, there is a global acceleration of the domain. Note, interestingly, that the sign of the third derivative of $a_\CD$ may also change, allowing for a variety of different dynamical evolution histories of the dark energy component. In the following, we pick two proto--type examples of instability sectors to which the dark components can effectively be associated.

Consider a small domain filled mainly with dust ($\Omega_m^\initial\CD \simeq 1$), dominated by its shear fluctuation ($\Omega_X^\initial\CD \gtrsim 0$), with an initially slightly positive constant curvature (\hbox{$\Omega_k^\initial\CD \lesssim 0$}) and such that $n \lesssim -3$. This situation can correspond to negative or positive averaged scalar curvature $\average{\CR}$, and it describes a pancake--like configuration of dust, like a proto--sheet (or proto--filament). Such a domain evolves toward a state dominated by the backreaction fluid, passing close to the observational ratio $\Omega_m^\CD/\Omega_X^\CD = \Omega_{baryon}/\Omega_{DM} \simeq 0.21$, and eventually collapses\footnote{Note that our effective model does not take into consideration either the pressure or the vorticity of the matter at stake, which obviously cannot be neglected at the end of the collapse process.} (see figures~\ref{fig:phase_portraits} and \ref{fig:scale_fact}). This example shows that the backreaction terms can turn out to play the role of a dark matter component, thereby accelerating the formation of structures by adding gravity and accreting the dust.

Consider a large domain, say our Hubble sphere, endowed with an initially slightly negative constant curvature ($\Omega_k^\initial\CD \gtrsim 0$) and mainly composed of dust ($\Omega_m^\initial\CD \simeq 1$). Suppose also that the expansion variance slowly varies and dominates the backreaction fluid (so $-2 < n \lesssim 0$ and $\Omega_X^\initial\CD \gtrsim 0$). Such a domain is always characterized by a negative averaged scalar curvature $\average{\CR}$ and enters, from a finite time onward, a phase of accelerated expansion, may pass close to the observational value $(\Omega_m^\CD, \Omega_X^\CD) \simeq (0.27, 0.73)$ (the square in figure~\ref{fig:phase_portraits}), and eventually ends up being entirely dominated by the backreaction fluid. Recall that the backreaction fluid can be formulated as a scalar field \cite{morphon}, and here its equation of state would be given by $\omega = -1 - n/3 \gtrsim -1$. These accelerated expansion scenarios include the case found in perturbation theory \cite{li_gaugeinv}, and they would be interpreted as the signature of dark energy in the standard model.

These two examples, though not depicting exhaustively all the possible scenarios, show how the instability sectors of FLRW backgrounds can be linked to the dark components.

\section{Concluding remarks} \label{sec:conc}

We have studied in the present paper the global gravitational stability of FLRW backgrounds in the class of scaling averaged inhomogeneous models. Our investigation has led to a detailed classification of stability and instability sectors in the phase space of the scaling averaged cosmologies, and it has shown that averaged models are driven away from FLRW backgrounds through structure formation and accelerated expansion. Furthermore, motivated by reasonable physical assumptions, we have been able to relate both dark components of the cosmological concordance model to the instability sectors of the dynamical system. 

As for the cases where the FLRW backgrounds are found to enjoy stability, we have to add a disclaimer. 
Despite their stable character, these FLRW backgrounds might not serve as reliable approximations in the role of physical backgrounds. As an example we look at the region II.b
(green) in figure 1: we appreciate that the background experiences a transient acceleration period in between an initial and an asymptotic Friedmann--like state. Such a situation  
shows that FLRW backgrounds, even if globally stable, do {\it a priori} not provide a correct approximation for the physical background during the entire evolution. 

A last remark concerns the use of exact scaling solutions for the class of averaged models analyzed. 
Indeed, there is no proof that scaling solutions depict all the instabilities of FLRW backgrounds, but it is also highly improbable that this is the only class of averaged models where instability occurs. Since our point was to explicitly demonstrate the possible unstable character of Friedmann--like domains, the use of scaling solutions is perfectly legitimate, though not exhaustive. In other words, our study is restrictive in this sense, since scaling solutions, albeit exact, are only indicative for the existing instabilities. One could also speculate that scaling solutions are acceptable building blocks to describe any backreaction behavior, as any analytical function is the sum of a Laurent series; however the correspondence is not that trivial due to the nonlinearity of the problem.

\medskip\noindent
{\it Acknowledgements:} 
{\small The authors are thankful for the valuable comments of the anonymous referees. XR thanks Miv Oto for useful discussions. This work is supported by ``F\'ed\'eration de Physique Andr\'e--Marie Amp\`ere'' of Universit\'e Lyon~1 and \'Ecole Normale Sup\'erieure de Lyon.}


{

\renewcommand{\theequation}{A.\arabic{equation}}
\setcounter{equation}{0}

\renewcommand{\thefigure}{A.\arabic{figure}}
\setcounter{figure}{0}


\end{document}